\documentclass{rstransa}

\begin{document}
\title{Collective in-plane magnetization in a 2D XY macrospin system within the framework of generalized Ott-Antonsen theory}


\author{
Irina V.\ Tyulkina$^{1}$, Denis S.\ Goldobin$^{1,2}$, Lyudmila S.\ Klimenko$^{1,2}$,
Igor S.\ Poperechny$^{1}$, and Yuriy L.\ Raikher$^{1}$}

\address{$^{1}$Institute of Continuous Media Mechanics, UB RAS, Academician Korolev Street 1, 614013 Perm, Russia\\
$^{2}$Department of Theoretical Physics, Perm State University, Bukirev Street 15, 614990 Perm, Russia}

\subject{Magnetism, Statistical Physics, Nonlinear Dynamics}

\corres{Yuriy L.\ Raikher\\
\email{yuriy.raikher@gmail.com}}

\begin{abstract}
The problem of magnetic transitions between the low-temperature (macrospin ordered) phases in 2D XY arrays is addressed.
The system is modeled as a plane structure of identical single-domain particles arranged in a square lattice and coupled by the magnetic dipole-dipole interaction;
all the particles possess a strong easy-plane magnetic anisotropy.
The basic state of the system in the considered temperature range is an antiferromagnetic (AF) stripe structure, where the macrospins (particle magnetic moments) are still involved in thermofluctuational motion: the superparamagnetic blocking $T_b$ temperature is lower than that ($T_\text{af}$) of the AF transition.
The description is based on the stochastic equations governing the dynamics of individual magnetic moments, where the interparticle interaction is added in the mean field approximation.
With the technique of a generalized Ott--Antonsen theory, the dynamics equations for the order parameters (including the macroscopic magnetization and the antiferromagnetic order parameter) and the partition function of the system are rigorously obtained and analysed.
We show that inside the temperature interval of existence of the AF phase, a static external field tilted to the plane of the array is able to induce first order phase transitions from AF to ferromagnetic state; the phase diagrams displaying stable and metastable regions of the system are presented.
\end{abstract}

\begin{fmtext}
\end{fmtext}

\maketitle

\section{Introduction  \label{sec:1}}
\subsection{General remarks}
Two-dimensional macrospin (2D XY) systems make a very rich object for modelling and investigation of the possible ordered states and the transitions between the latter.
These systems are well reproduced experimentally in the form of plane arrays of nanodisks of 10 to 100 nm size divided by the gaps of the same order of magnitude.
Due to that, well below the the Curie temperature of the ferromagnet the disks are made of, they behave as classical single-domain particles with magnetic anisotropy of easy-plane type originating from their flat shape, see \cite{Arnalds-etal-2014,Leo-etal-2018}, for example.
Under such a large spatial separation, the disks are completely free from the exchange interaction.
In this situation, the two factors, which affect the individual and collective response of the magnetic moments, are: thermal fluctuations and dipole-dipole interaction.

This combination imparts to 2D XY ensembles some remarkable properties.
The most interesting of those is their fundamental ability to form magnetically ordered states at the temperatures above those where the actual orientations of the particle magnetic moments are fixed by the superparamagnetic blockade.
Moreover, the orientational thermal fluctuations of the particle magnetic moments are the necessary condition for self-organisation of the magnetically ordered states --- antiferromagnetic (AF) and ferromagnetic (FM) --- in 2D XY ensembles.

When building up the statistical thermodynamics of these systems, one comes up to the ubiquitous difficulty: the necessity to evaluate the partition function for a multiparticle ensembles with strong long-range interaction.
For the case of 2D XY we have found a way to do that, although in the mean field approximation, but otherwise rigorously, using the technique of circular cumulants~\cite{Tyulkina-etal-2018} that is a generalisation of the Ott--Antonsen theory~\cite{Ott-Antonsen-2008,Ott-Antonsen-2009}.

\subsection{Opportunities of the Ott--Antonsen theory and its generalization}
In the theory of collective phenomena, many paradigmatic models \cite{Kuramoto-2003,Pikovsky-Rosenblum-Kurths-2003,Dietert-Fernandez-2018} are governed by equations of the form
\begin{equation}\label{eq:101}
\dot\varphi_j=\omega(t)+\mathrm{Im}(2\mathcal{H}(t)e^{-i\varphi_j})\,, \qquad j=1,...,N,
\end{equation}
\noindent
where variables $\varphi_j$ are either the angles of directional elements or do characterize the oscillation phase for elements with periodic self-oscillations.
Here real-valued $\omega(t)$ and complex-valued $\mathcal{H}(t)$ can be functions of time and ensemble state $\{\varphi_l|l=1,2,...,N\}$ of arbitrary complexity; it is only important that they are identical for all elements.
Chains of superconducting Josephson junctions were the first systems of this sort, for which the peculiar mathematical properties allowed one to characterize the collective dynamics in great detail \cite{Watanabe-Strogatz-1993,Watanabe-Strogatz-1994}.
Finally, the Watanabe--Strogatz theory \cite{Watanabe-Strogatz-1993,Watanabe-Strogatz-1994,Pikovsky-Rosenblum-2008,Marvel-Mirollo-Strogatz-2009} was developed for ensembles (\ref{eq:101}) with finite $N\ge3$.
This theory established a foundation for the Ott--Antonsen (OA) theory \cite{Ott-Antonsen-2008,Ott-Antonsen-2009}, which yields a closed equation for the dynamics of the order parameter $Z=N^{-1}\sum_{j=1}^Ne^{i\varphi_j}$ in the thermodynamic limit $N\to\infty$\,:
\begin{equation*}
\dot{Z}=i\omega(t)Z+\mathcal{H}(t)-\mathcal{H}^\ast(t)\,Z^2.
\end{equation*}

The opportunity to have an exact closed equation for the dynamics of the order parameter resulted in an eruption of works employing this mathematical tool.
The Ott--Antonsen theory proved itself to be a useful tool for studies on Josephson junction arrays \cite{Marvel-Strogatz-2009}, neuronal networks \cite{Laing-2014,Pazo-Montbrio-2014,Montbrio-Pazo-Roxin-2015,Luke-Barreto-So-2014,Laing-2018b}, populations of active rotators \cite{Dolmatova-etal-2017,Klinshov-Franovic-2019}, fundamental studies on collective phenomena \cite{Abrams-etal-2008,Laing-2015,Laing-2009,Bordugov-Pikovsky-Rosenblum-2010,Omelchenko-etal-2014,Smirnov-etal-2017,Smirnov-etal-2018,Omelchenko-2019,Laing-2017,Omelchenko-2018,Pimenova-etal-2016,Goldobin-etal-2017}, etc.
Below we will show that the in-plane dynamics of magnetic moments in a system, where the interaction between elements is mediated by the magnetic field \cite{Arnalds-etal-2014,Leo-etal-2018}, is also governed by equations identical or similar to (\ref{eq:101}).
Nonetheless, the original OA theory \cite{Ott-Antonsen-2008,Ott-Antonsen-2009} could be only of very limited use for collective magnetism problems, since it cannot handle the thermal noise and can deal with just particular types of nonidentities of parameters of individual elements.

In real systems, the form of equations (\ref{eq:101}) is obviously distorted (see \cite{Kuramoto-Nakao-2019,Ermentrout-Park-Wilson-2019}, for example), and the generalization of the OA theory to nonideal situations was a resisting challenge for a decade.
A way out has been proposed recently in the form of circular cumulant approach \cite{Tyulkina-etal-2018,Tyulkina-etal-2019,Goldobin-Dolmatova-PRR-2019,Goldobin-2019}.
This technique allows one to generalize the OA theory and derive closed equation systems for the dynamics of order parameters in the presence of thermal noise (\textit{or} `intrinsic noise') and under other violations of the applicability conditions of the original OA theory.

In what follows we show in detail how a generalized version of the Ott--Antonsen theory could be applied for macroscopic description of the collective magnetism phenomena in systems with one principal angular degree of freedom, namely XY macrospin systems.
We first consider the individual dynamics of magnetic moments of strongly magnetically anisotropic spheroidal single-domain magnetic particles arranged in a square 2D array on the plane \cite{Arnalds-etal-2014,Leo-etal-2018}.
The magnetic moments are subject to thermal fluctuations and experience the magnetic friction, in accordance with the fluctuation-dissipation theorem.
For this system, taking the Landau-Lifshitz-Gilbert magnetodynamic equation as a starting point, in section \ref{sec:2} an approximate equation of the azimuthal angle dynamics is derived.
In section \ref{sec:3} a brief introduction to the OA theory and its generalization is given.
In section \ref{sec:4}, on the basis of this generalization, a closed set of governing equations for the order parameters of the magnetic sublattices are written down and the specific terms of these equations for the case of antiferromagnetic states in the presence of an external magnetic field are obtained.
Analytical and numerical study of these equations are given in section \ref{sec:5} together with description of sthe phase transitions between antiferromagnetic, ferromagnetic, and paramagnetic macroscopic states in the array.
The evidence obtained is summarised in section \ref{sec:6}.

\section{Dynamics of the magnetic moment of a single-domain\\ ferromagnetic particle subject to thermal noise  \label{sec:2}}
\subsection{Spheroidal magnetic particle in quasistatic magnetic field}
The dynamics of the magnetic moment of an immobilized single-domain spheroidal particle is governed by the Landau--Lifshitz--Gilbert \cite{LaLi-PZS:35,fmr,Gilbert-2004-1955} equation
\begin{equation}\label{eq:201}
\frac{\mathrm{d}\vec{M}}{\mathrm{d}t}=-\gamma\vec{M}\times\vec{H}_\mathrm{eff} +\frac{\eta}{M}\vec{M}\times\frac{\mathrm{d}\vec{M}}{\mathrm{d}t}
 +\widetilde{\sigma}\vec{\zeta}(t)\,,
\end{equation}
where the magnitude of the magnetic moment $M$ is constant, $-\gamma$ is the gyromagnetic ratio, $\eta$ is the dimensionless magnetic relaxation coefficient, $\widetilde{\sigma}$ is the strength of thermal noise, $\vec{\zeta}(t)$ is the normalised white vector Gaussian noise; $\vec{\zeta}\perp\vec{M}$ and its mutually orthogonal components are independent. Below, we will specify the properties of thermal noise term in more detail, with account for the fluctuation--dissipation theorem \cite{Callen-Welton-1951,Einstein-1905}. For a spheroidal particle with a magnetic easy $(x,y)$-plane, potential energy $U=-\vec{M}\cdot\vec{H}+\frac{1}{2}KV(\vec{e}_M\cdot\vec{e}_{z})^{2}$ and
\begin{equation*}
\vec{H}_\mathrm{eff}=-\frac{\partial U}{\partial\vec{M}}=\vec{H}-\frac{KV}{M^2}M_{z}\vec{e}_{z}\,,
\end{equation*}
where $\vec{H}$ is an external field, $V$ is the particle volume, $KV/2$ is the magnetic anisotropy energy, $\vec{e}_M$ and $\vec{e}_z$ are the unit vectors along the magnetic moment and the $z$-axis.

\begin{figure}[!h]
\centering
\textit{(a)}\includegraphics[height=1.25in]{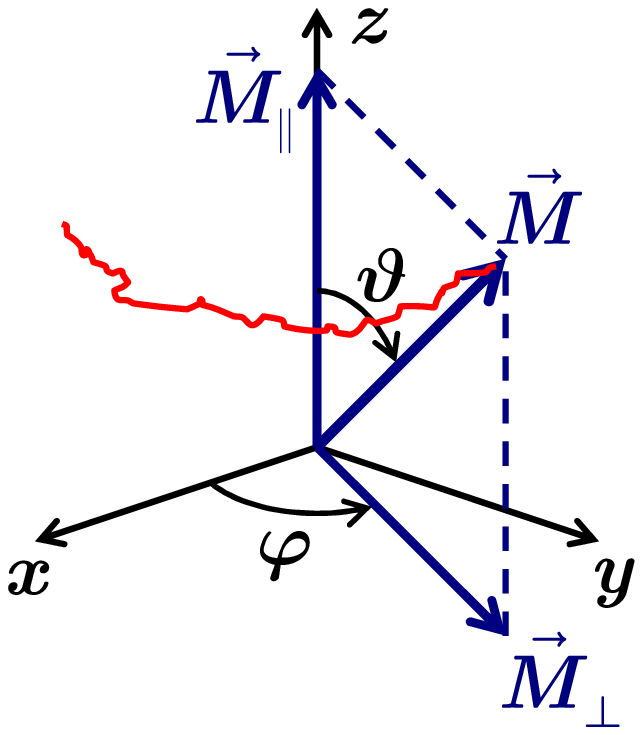}
\hspace{0.75in}
\textit{(b)}\includegraphics[height=1.4in]{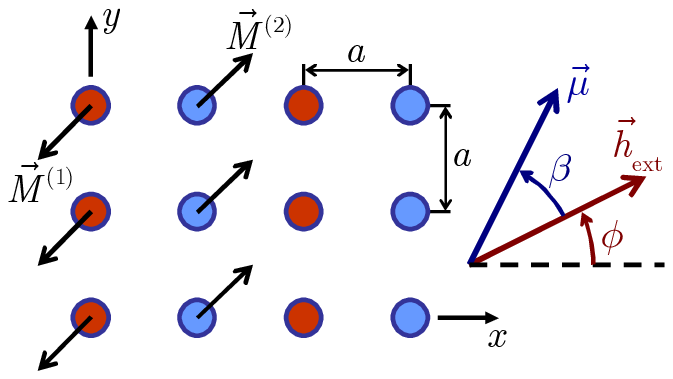}
\caption{\textit{(a)} Coordinate frame and trajectory of an individual magnetic moment. \textit{(b)}~XY spin system with two magnetic sublattices on a square array of single-domain magnetic particles.}
\label{fig1}
\end{figure}

For convenience, we decompose the magnetic field $\vec{H}=\vec{H}_0+\vec{h}$ into the $z$-component $\vec{H}_0=H_0\vec{e}_z$ and the $(x,y)$-component $\vec{h}=h_{0}\{\cos\Phi,\sin\Phi,0\}$.
In the consideration below, we will admit $H_0$, $h_0$, $\Phi$ to vary with time slowly; the criterion for the ``slow'' dynamics will be also specified below.
Let us recast (\ref{eq:201}) in the spherical coordinate frame (see figure \ref{fig1}\textit{a}),
\begin{equation*}
\vec{M}=M\{\sin\vartheta\cos\varphi, \sin\vartheta\sin\varphi, \cos\vartheta\}.
\end{equation*}
In spherical coordinates, equation (\ref{eq:201}) reads
\begin{equation*}
\left\{
\begin{split}
\dot{\vartheta}+\eta\sin\vartheta\dot{\varphi}&=\gamma h_{0}\sin(\Phi-\varphi)+\frac{\tilde{\sigma}}{M}\zeta_\vartheta(t)\,,
\\
-\eta\dot{\vartheta}+\sin\vartheta\dot{\varphi}&=-\gamma h_{0}\cos\vartheta\cos(\Phi-\varphi) +\gamma\left(H_{0}-\frac{KV}{M}\cos\vartheta\right)\sin\vartheta
+\frac{\tilde{\sigma}}{M}\zeta_\varphi(t)\,,
\end{split}
\right.
\end{equation*}
where $\zeta_\vartheta$ and $\zeta_\varphi$ are the components of thermal noise in the polar angle and azimuthal directions, respectively. From the latter equation system, one can obtain

\begin{equation}\label{eq:202}
\left\{
\begin{split}
\dot{\vartheta}&=\frac{\gamma h_{0}}{1+\eta^2}\left[\sin(\Phi-\varphi)+\eta\cos\vartheta\cos(\Phi-\varphi)\right]
-\frac{\eta\gamma}{1+\eta^2}(H_{0}-{\textstyle\frac{KV}{M}}\cos\vartheta)\sin\vartheta +\sigma\zeta_{1}(t)\,,
\\
\dot{\varphi}&=\frac{\gamma h_{0}}{1+\eta^2}\frac{1}{\sin\vartheta}\left[\eta\sin(\Phi-\varphi)-\cos\vartheta\cos(\Phi-\varphi)\right]
+\frac{\gamma}{1+\eta^2}(H_{0}-{\textstyle\frac{KV}{M}}\cos\vartheta)+\frac{\sigma\zeta_{2}(t)}{\sin\vartheta}\,,
\end{split}
\right.
\end{equation}
where $\zeta_1(t)\equiv\frac{\zeta_\vartheta(t)-\eta\zeta_\varphi(t)}{\sqrt{1+\eta^2}}$, $\zeta_2(t)\equiv\frac{\zeta_\varphi(t)+\eta\zeta_\vartheta(t)}{\sqrt{1+\eta^2}}$ and $\sigma\equiv\frac{\widetilde\sigma}{M\sqrt{1+\eta^2}}$.

With
$\langle\zeta_\vartheta(t)\,\zeta_\vartheta(t^\prime)\rangle=\langle\zeta_\varphi(t)\,\zeta_\varphi(t^\prime)\rangle=2\delta(t-t^\prime)$ and
$\langle\zeta_\vartheta(t)\,\zeta_\varphi(t^\prime)\rangle=0$, where $\langle\dots\rangle$ indicates the averaging over noise realizations,
one can calculate
\begin{equation*}
\begin{split}
\langle\zeta_1(t)\,\zeta_1(t^\prime)\rangle& =\left\langle\frac{(\zeta_\vartheta(t)-\eta\zeta_\varphi(t)) (\zeta_\vartheta(t^\prime)-\eta\zeta_\varphi(t^\prime))}{1+\eta^2}\right\rangle
=2\delta(t-t^\prime)\,,
\\
\langle\zeta_2(t)\,\zeta_2(t^\prime)\rangle& =\left\langle\frac{(\zeta_\varphi(t)+\eta\zeta_\vartheta(t)) (\zeta_\varphi(t^\prime)+\eta\zeta_\vartheta(t^\prime))}{1+\eta^2}\right\rangle=2\delta(t-t^\prime)\,,
\\
\langle\zeta_1(t)\,\zeta_2(t^\prime)\rangle&
=\left\langle\frac{(\zeta_\vartheta(t)-\eta\zeta_\varphi(t)) (\zeta_\varphi(t^\prime)+\eta\zeta_\vartheta(t^\prime))}{1+\eta^2}\right\rangle=0\,.
\end{split}
\end{equation*}
Hence, $\zeta_1(t)$ and $\zeta_2(t)$  are mutually independent normalized $\delta$-correlated noise signals. For $h_0=0$ and constant $H_0$, one can evaluate the distribution of $\vartheta$ from~(\ref{eq:202}); this distribution will coincide with the thermodynamic equilibrium distribution $w(\vartheta)=const\,\sin\vartheta e^{-\frac{U(\vartheta)}{kT}}$ if
\begin{equation}\label{eq:203}
\sigma^2=\frac{kT\eta\gamma}{(1+\eta^2)M}\,,
\end{equation}
where $k$ is the Boltzmann constant and $T$ is temperature.
As the thermal fluctuation intensity does not depend on external field $\vec{h}$, the thermal noise intensity dictated by the fluctuation--dissipation theorem is given by (\ref{eq:203}).

\subsection{Azimuthal angle reduction of the dynamics of magnetic moment}
In what follows, we consider the case of strong external magnetic field $H_0$ and large anisotropy energy; we assume the field $H_0$ to be not sufficiently strong to overcome the magnetic anisotropy and align the magnetic moment along the $z$-axis:
\begin{equation}\label{eq:204}
MH_0\sim KV\gg Mh_0\sim \sigma^2M/\gamma\,,\qquad MH_0<KV\,.
\end{equation}
In this case, the in-plane dynamics of the magnetic moment is relatively slow and the relaxation of the perpendicular to the plane magnetization is fast; therefore, a reduction of the dynamics dimensionality should be possible due to the separation of time scales~\cite{Fenichel-1979}.

According to the first equation of system (\ref{eq:202}), the polar angle $\vartheta$ fluctuates within a small vicinity of $\vartheta_\ast$ determined by the condition $KV\cos\vartheta_\ast=MH_0$;
\begin{equation*}
\vartheta=\vartheta_\ast+\vartheta_1,\qquad \cos\vartheta_\ast=\frac{MH_0}{KV},\quad |\vartheta_1|\ll1\,.
\end{equation*}

When $h_0$, $(\Phi-\varphi)$, and $H_0$ evolve slowly compared to the relaxation rate of fluctuations $\vartheta_1$
\begin{equation}\label{eq:205}
\lambda\equiv\frac{\eta\gamma}{1+\eta^2}\frac{KV}{M}\sin^2\vartheta_\ast=\frac{\eta\gamma}{1+\eta^2}\frac{KV}{M}\left(1-\frac{(MH_0)^2}{(KV)^2}\right),
\end{equation}
the linear in $\vartheta_1$ approximation of the first equation of~(\ref{eq:202}) yields
\[
\vartheta_1\approx\frac{Mh_{0}\left[\sin(\Phi-\varphi) +\eta\cos\vartheta_{*}\cos(\Phi-\varphi)\right]}{\eta KV\Big(1-\frac{(MH_0)^2}{(KV)^2}\Big)}
+\sigma\int_{0}^{\infty}\mathrm{d}\tau\,\zeta_{1}(t-\tau)\,e^{-\lambda\tau}.
\]
Hence, to the linear in $\vartheta_1$ terms, the second equation of (\ref{eq:202}) yields for the azimuthal angle
\begin{align}
&\dot{\varphi} \approx\frac{\gamma h_{0}}{1+\eta^2}\frac{\eta\sin(\Phi-\varphi)-\cos\vartheta_\ast\cos(\Phi-\varphi)}{\sin\vartheta_\ast}
+\frac{\gamma h_{0}\sin\vartheta_\ast}{1+\eta^2}\frac{\sin(\Phi-\varphi)+\eta\cos\vartheta_\ast\cos(\Phi-\varphi)} {\eta\Big(1-\frac{(MH_0)^2}{(KV)^2}\Big)}
\nonumber\\
&\qquad
+\frac{\gamma}{1+\eta^2}\frac{KV}{M}\sin\vartheta_\ast\sigma\int_{0}^{\infty}\mathrm{d}\tau\,\zeta_1(t-\tau)\,e^{-\lambda\tau} +\frac{\sigma\zeta_2(t)}{\sin\vartheta_\ast}\,.
\label{eq:206}
\end{align}

In the limit $\lambda\to\infty$, the signal $\xi_1(t)=\lambda\int_{0}^{\infty}\zeta_1(t-\tau)e^{-\lambda\tau}\mathrm{d}\tau$ in (\ref{eq:206}) becomes a $\delta$-correlated noise: its autocorrelation function for $\tau>0$
\begin{align}
&\langle\xi_1(t)\,\xi_1(t+\tau)\rangle=\lambda^2\left\langle\int_0^\infty\mathrm{d}\tau_1\zeta_1(t-\tau_1)\,e^{-\lambda\tau_1}
\int_0^\infty\mathrm{d}\tau_2\zeta_1(t+\tau-\tau_2)\,e^{-\lambda\tau_2}\right\rangle \nonumber \\
&\qquad=\lambda^2\int_0^\infty\mathrm{d}\tau_1 \int_0^\infty\mathrm{d}\tau_2e^{-\lambda(\tau_1+\tau_2)}2\delta(\tau_1-\tau_2+\tau)
=2\lambda^2\int_{0}^{\infty}\mathrm{d}\tau_{1}e^{-\lambda(2\tau_{1}+\tau)}
=\lambda e^{-\lambda|\tau|}. \nonumber
\end{align}
As $\int_{-\infty}^{+\infty}\lambda e^{-\lambda\vert\tau\vert}\mathrm{d}\tau=2$, the limit $\lim_{\lambda\to0}\langle\xi_1(t)\,\xi_1(t+\tau)\rangle=2\delta(\tau)$, i.e., $\xi_1(t)$ is normalized.
The sum of two independent $\delta$-correlated Gaussian noises is a $\delta$-correlated Gaussian noise; the Gaussian noise intensities are additive:
\begin{equation*}
\frac{\sigma\gamma KV\sin\vartheta_\ast}{(1+\eta^2)M\lambda}\xi_1(t) +\frac{\sigma\zeta_2(t)}{\sin\vartheta_\ast}\to
\sqrt{\left(\frac{\sigma\gamma KV\sin\vartheta_\ast}{(1+\eta^2)M\lambda}\right)^2 +\frac{\sigma^2}{\sin^2\vartheta_\ast}}\zeta_3(t) =\frac{\sigma\zeta_3(t)}{\sin\vartheta_\ast}\sqrt{\frac{1}{\eta^2}+1}\,,
\end{equation*}
where $\zeta_3(t)$ is a normalized $\delta$-correlated Gaussian noise.

Thus, for the case~(\ref{eq:204}), which by virtue of relation~(\ref{eq:205}) results also in $\lambda\gg\gamma h_0$, the magnetic moment dynamics reduces to a one-angle stochastic dynamics with a single additive effective noise term; equation~(\ref{eq:206}) with (\ref{eq:203}) yields
\begin{equation}\label{eq:207}
\dot{\varphi}=\frac{KV}{\sqrt{K^2V^2-M^2H_0^2}}\left(\frac{\gamma h_{0}}{\eta}\sin(\Phi-\varphi)+\sqrt{\frac{\gamma kT}{\eta M}}\,\zeta_3(t)\right).
\end{equation}
The multiplier ahead of the brackets is an expression for $1/\sin\vartheta_\ast$\,.

\section{Generalized Ott--Antonsen theory and macroscopic magnetization  \label{sec:3}}
Here we give a brief introduction to the Ott--Antonsen theory and its generalization in terms of circular cumulants. Basically, the former is formulated for an ensemble of identical phase/angle elements governed by equations
\begin{equation}\label{eq:OA1}
\dot\varphi_j=\omega(t)+\mathrm{Im}(2\mathcal{H}(t)e^{-i\varphi_j})\,,
\qquad j=1,...,N,
\end{equation}
where $\omega(t)$ and $\mathcal{H}(t)$ are arbitrary real- and complex-valued functions of time; $N$ is the ensemble size. With regard to the physical system we consider, one can notice that equation~(\ref{eq:207}) with $\sigma=0$ corresponds to (\ref{eq:OA1}) with
 $\mathcal{H}=\gamma h_0e^{i\Phi}/(2\eta\sin\vartheta_\ast)$ and $\omega=0$.
The OA theory is valid in the thermodynamic limit $N\to\infty$, where the system state is naturally represented by the probability density function $w(\varphi,t)$. The master equation for $w(\varphi,t)$ reads
\begin{equation}\label{eq:OA2}
\frac{\partial w}{\partial t}+\frac{\partial}{\partial\varphi}\left[\left(\omega(t) -i\mathcal{H}(t)e^{-i\varphi}+i\mathcal{H}^\ast(t)e^{i\varphi}\right)w\right]=0\,.
\end{equation}
In Fourier space, where
\begin{equation}\label{eq:OA3}
w(\varphi,t)=\frac{1}{2\pi}\Big[1+\sum_{m=1}^\infty \left(Z_m(t)e^{-im\varphi}+Z_m^\ast(t)e^{im\varphi}\right)\Big]
\end{equation}
and $Z_m(t)=\int_{0}^{2\pi}w(\varphi,t)e^{im\varphi}\mathrm{d}\varphi=\langle{e^{im\varphi}}\rangle$,
master equation (\ref{eq:OA2}) takes the form
\begin{equation}\label{eq:OA4}
\dot{Z}_m=im\omega Z_m+m\mathcal{H}Z_{m-1}-m\mathcal{H}^\ast{Z}_{m+1}\,,
\end{equation}
where $Z_0=1$ and $Z_{-m}=Z_m^\ast$ by definition. Ott and Antonsen \cite{Ott-Antonsen-2008} noticed that equation system~(\ref{eq:OA4}) admits solution $Z_m(t)=\left[Z_1(t)\right]^m$ with order parameter $Z_1=\langle{e^{i\varphi}}\rangle$ governed by a simple self-contained equation:
\begin{equation}\label{eq:OA5}
\dot{Z}_1=i\omega Z_1+\mathcal{H}-\mathcal{H}^\ast{Z}_1^2\,.
\end{equation}
In the literature, the substitution $Z_m=\left(Z_1\right)^m$ is referenced to as the Ott--Antonsen ansatz.

In the case of an ensemble of magnetic moments, the planar component of the mean moment can be characterized by $Z_1$:
\begin{equation}\label{eq:OA6}
\langle{M_x}\rangle+i\langle{M_y}\rangle=M\sin\vartheta_\ast\langle{e^{i\varphi}}\rangle=M\sin\vartheta_\ast Z_1\,.
\end{equation}
Thus, a self-contained equation for the dynamics of $Z_1$ provides a detailed characterization of the macroscopic magnetization of an ensemble.

The issue of the attractivity of the discovered particular solution was also addressed in the OA theory.
The OA manifold $Z_m=\left(Z_1\right)^m$ is neutrally stable for perfectly identical population elements, but becomes attracting for typical cases of imperfect parameter identity, where the parameter distribution is continuous \cite{Ott-Antonsen-2009,Mirollo-2012,Pietras-Daffertshofer-2016}, or in the presence of weak additive intrinsic noise \cite{Tyulkina-etal-2018} --- the thermal noise in our case. Thus, this solution is attracting for real situations, which are always imperfect, and is of practical interest. Equation~(\ref{eq:OA5}) is an exact result, which provides a closed equation for the dynamics of order parameter $Z_1$ and made a ground for a significant advance in various studies on collective phenomena.

For applications of the OA theory it was important to be able to deal with ensembles of non-identical elements, since the ensembles of identical elements~(\ref{eq:OA1}) typically tend to perfect order $|Z_1|=1$ or maximal disorder $Z_1=0$ (notice, no thermal noise in (\ref{eq:OA1})). For the cases of a Lorentzian distribution or other fractional rational distributions of $\omega_j$, one can rigorously derive a modified version of (\ref{eq:OA5}) \cite{Ott-Antonsen-2008,Ott-Antonsen-2009,Pietras-Daffertshofer-2016} and study the imperfect order states in great detail. This approach is not limited to the cases of nonidentity of $\omega_j$ and can be applied for the cases of nonidentity of some coefficients in $\mathcal{H}(t)$ \cite{Pazo-Montbrio-2014,Montbrio-Pazo-Roxin-2015}. However, for the case of magnetic moment ensemble we consider, these forms of nonidentity are not relevant. Whilst the case of thermal noise cannot be handled within the framework of the original OA theory.

For the XY spin system we consider, we need a generalization of the OA theory for the case of ensemble of identical elements with individual intrinsic noise:
\begin{equation}\label{eq:OA7}
\dot\varphi_j=\omega(t)+\mathrm{Im}(2\mathcal{H}(t)e^{-i\varphi_j})+\sqrt{D}\zeta_j(t)\,, \qquad j=1,...,N,
\end{equation}
where
\begin{equation}\label{eq:defD}
D=\frac{\gamma kT}{\eta M\sin^2\vartheta_\ast}
\end{equation}
is the noise intensity, $\zeta_j(t)$ are independent normalized Gaussian noise signals:
 $\langle\zeta_j(t)\rangle=0$, $\langle\zeta_j(t)\,\zeta_l(t^\prime)\rangle=2\delta_{jl}\delta(t-t^\prime)$, and $\delta_{jl}$ is the Kronecker delta which is $1$ for $j=l$ and $0$ otherwise.
In the presence of thermal noise, master equation~(\ref{eq:OA2}) turns into the Fokker--Planck equation
\begin{equation}\label{eq:OA8}
\frac{\partial w}{\partial t} +\frac{\partial}{\partial\varphi}\left[\left(\omega(t) -i\mathcal{H}(t)e^{-i\varphi}+i\mathcal{H}^\ast(t)e^{i\varphi}\right)w\right] -D\frac{\partial^2w}{\partial\varphi^2}=0\,,
\end{equation}\label{eq:OA9}
which yields in Fourier space, instead of~(\ref{eq:OA4}),
\begin{equation}
\dot{Z}_m=im\omega Z_m+m\mathcal{H}Z_{m-1}-m\mathcal{H}^\ast{Z}_{m+1}-m^2DZ_m\,.
\end{equation}
The latter equation system does not admit the OA ansatz $Z_m=(Z_1)^m$.
In \cite{Tyulkina-etal-2018,Goldobin-Dolmatova-PRR-2019}, a circular cumulant approach was developed for tackling the collective behavior of ensembles beyond the OA ansatz and, in particular, dealing with equation system~(\ref{eq:OA9}).

Let us consider $Z_m$ as moments of $e^{i\varphi}$ and formally introduce corresponding cumulants~\cite{Tyulkina-etal-2018}. The latter quantities are not conventional cumulants of original variable $\varphi$; therefore, we are free to choose the normalization for them and refer to them as `circular cumulants'. With the moment generating function
\begin{equation*}
F(k)\equiv\langle\exp(ke^{i\varphi})\rangle=1+Z_1k+Z_2\frac{k^2}{2!}+Z_3\frac{k^3}{3!}+\dots
\end{equation*}
we define circular cumulants $\kappa_m$ via the generating function
\begin{equation*}
\Psi(k)\equiv k\frac{\partial}{\partial k}\ln{F(k)}\equiv\kappa_1k+\kappa_2k^2+\kappa_3k^3+\dots\,.
\end{equation*}
For example, the first three circular cumulants are
\begin{equation*}
\kappa_1=Z_1, \qquad \kappa_2=Z_2-Z_1^2, \qquad \kappa_3=(Z_3-3Z_2Z_1+2Z_1^3)/2.
\end{equation*}

In terms of circular cumulants, the OA manifold $Z_m=(Z_1)^m$ acquires a simple form:
\begin{equation*}
\kappa_1=Z_1\,,\qquad \kappa_{m\ge2}=0\,.
\end{equation*}
Thus, the Ott--Antonsen ansatz can be considered as the one-cumulant truncation of a circular cumulant series.

In terms of $\kappa_m$, equation system (\ref{eq:OA9}) turns into
\begin{align}
\dot{\kappa}_m=im\omega\kappa_m+\mathcal{H}\delta_{1m} -\mathcal{H}^\ast\Big(m^2\kappa_{m+1}+m\sum_{j=1}^m\kappa_{m-j+1}\kappa_j\Big) \nonumber\\
-D\Big(m^2\kappa_m+m\sum_{j=1}^{m-1}\kappa_{m-j}\kappa_j\Big)
\label{eq:OA10}
\end{align}
(see \cite{Tyulkina-etal-2018} for the regular derivation procedure). Although the latter equation system is more lengthy than equation system (\ref{eq:OA9}) for $Z_m$, it is much more convenient for dealing with. First, it is free of the loss of convergence for highly ordered states where $|Z_m|\to1$. Second, it is convenient for constructing perturbation theories, as $\kappa_m$ form a decaying geometric progression --- a hierarchy of smallness appears~\cite{Tyulkina-etal-2018,Goldobin-Dolmatova-PRR-2019}; in particular, $\kappa_m\propto D^{m-1}$ for $D\ll|\mathcal{H}|$, and $\kappa_m\propto(1/D)^m$ for $D\gg|\mathcal{H}|$.

In \cite{Tyulkina-etal-2018,Goldobin-etal-2018,Tyulkina-etal-2019,Goldobin-Dolmatova-PRR-2019,Ratas-Pyragas-2019,Goldobin-Dolmatova-2019b},
the circular cumulant approach was reported to be an efficient tool for studying the population dynamics beyond the OA ansatz. To have a leading order correction to the OA dynamics, one need to include $\kappa_2$ into consideration; equation system~(\ref{eq:OA10}) for $m=1,2$ yields
\begin{equation}\label{eq:OA11}
\begin{array}{l}
\dot{Z}_1=i\omega Z_1+\mathcal{H}-\mathcal{H}^\ast(Z_1^2+\kappa_2)-DZ_1\,, \\[5pt]
\dot{\kappa}_2=i2\omega\kappa_2-4\mathcal{H}^\ast(\kappa_3+Z_1\kappa_2) -D(4\kappa_2+2Z_1^2)\,.
\end{array}
\end{equation}
To make this equation system self-contained, one has to adopt some assumption on $\kappa_3$. The hierarchy of smallness of $\kappa_m$ emerging in system~(\ref{eq:OA10}) suggests the simplest closure $\kappa_3=0$. In~\cite{Goldobin-etal-2018}, it was shown that two-cumulant truncation~(\ref{eq:OA11}) with $\kappa_3=0$ yields approximate solutions the relative error of which rarely reaches $1\%$ and often stays several orders of magnitude bellow this level.

Physically, order parameter $Z_1$ represents the dipole mode of the orientational distribution of $\vec{M}_\perp$. Without thermal noise, angle $\varphi$ obeys the distribution with $Z_m=(Z_1)^m$, which is a wrapped Cauchy (Lorentzian) distribution~\cite{Ley-Verdebout-2017,Goldobin-Dolmatova-PRR-2019}:
\begin{equation*}
w_\mathrm{OA}(\varphi)=\sum_{n=-\infty}^{+\infty}\frac{\pi^{-1}\ln|Z_1|}{\ln^2|Z_1|+(\varphi-\arg{Z_1}+2\pi n)^2}
 =\frac{1}{2\pi}\frac{1-|Z_1|^2}{1-|Z_1|\cos(\varphi-\arg{Z_1})}\,.
\end{equation*}
This distribution is controlled by $Z_1$. The second cumulant $\kappa_2=Z_2-Z_1^2$ quantifies the deviation of the quadrupole mode $Z_2$ of the distribution from the value $Z_1^2$ dictated by the wrapped Cauchy distribution ({\it or} the Ott--Antonsen solution).

\section{Two-sublattice mean-field theory for XY spin system  \label{sec:4}}
The analysis of the previous sections concerns the behavior of the mean magnetization for a superimposed magnetic field at a given location. Below we specify this magnetic field: it is assumed to be the superposition of a time-independent external magnetic field and the magnetic field from magnetic particles arranged into the 2D square array laying in the $(x,y)$-plane with the side size $a$ (see figure~\ref{fig1}\textit{b}). Without an external magnetic field, in such systems, two sorts of the local minima of potential energy exist: ferromagnetic and antiferromagnetic states. For the antiferromagnetic states, the minimal energy is achieved for the sublattices forming parallel stripes (figure~\ref{fig1}\textit{b}). Hence, we introduce two sublattices and will describe the dynamics of the order parameters for each sublattice.

We construct the mean-field theory, where the formation of domains with different mean magnetization is discarded from consideration. The magnetic field from each sublattice is approximately calculated as the magnetic field from the mean magnetic moments arranged into the corresponding spatial array. Mathematically, this means that the ensemble states are statistically homogeneous in space.

\subsection{Magnetic fields from sublattices}
Let us consider the planar component of the magnetic field acting on the node of sublattice $1$ from the same sublattice $1$ and sublattice $2$:
\begin{align}
\vec{H}^{(1\to1)}&
=\sum\limits_{j=-\infty}^{+\infty}\sum\limits_{l=-\infty}^{+\infty} \left(\frac{3(\vec{r}_{2j,l}\cdot\langle\vec{M}_{\perp}^{(1)}\rangle)\vec{r}_{2j,l}}{(r_{2j,l})^5} -\frac{\langle\vec{M}_{\perp}^{(1)}\rangle}{(r_{2j,l})^3}\right)\equiv\hat{\mathcal{L}}^{(1)}\cdot\frac{\langle\vec{M}_{\perp}^{(1)}\rangle}{a^3}\,, \nonumber\\
\vec{H}^{(2\to1)}&
=\sum\limits_{j=-\infty}^{+\infty}\sum\limits_{l=-\infty}^{+\infty} \left(\frac{3(\vec{r}_{2j-1,l}\cdot\langle\vec{M}_{\perp}^{(2)}\rangle)\vec{r}_{2j-1,l}}{(r_{2j-1,l})^5} -\frac{\langle\vec{M}_{\perp}^{(2)}\rangle}{(r_{2j-1,l})^3}\right)\equiv\hat{\mathcal{L}}^{(2)}\cdot\frac{\langle\vec{M}_{\perp}^{(2)}\rangle}{a^3}\,, \nonumber
\end{align}
where $\vec{M}_\perp^{(n)}=\{M_x^{(n)},M_y^{(n)},0\}$,  $\vec{r}_{j,l}=\{ja,la,0\}$ (see figure~\ref{fig1}\textit{b}), and matrices $\hat{\mathcal{L}}^{(1)}$ and $\hat{\mathcal{L}}^{(2)}$ are diagonal~\cite{Smart-1966,Vonsovskii-1974}:
\begin{equation}\label{eq:301}
\hat{\mathcal{L}}^{(1)}=\left(
\begin{array}{cc}
  -0.75876... & 0 \\
  0 & 4.80783...
\end{array}
\right)\,,
\qquad
\hat{\mathcal{L}}^{(2)}=\left(
\begin{array}{cc}
  5.27557... & 0 \\
  0 & -0.29104...
\end{array}
\right)\,.
\end{equation}
Hence, the magnetic field acting on sublattice $1$
\begin{equation*}
\vec{h}^{(1)}=\vec{h}_\mathrm{ext} +\hat{\mathcal{L}}^{(1)}\cdot\frac{\langle\vec{M}_{\perp}^{(1)}\rangle}{a^3} +\hat{\mathcal{L}}^{(2)}\cdot\frac{\langle\vec{M}_{\perp}^{(2)}\rangle}{a^3},
\end{equation*}
where $\vec{h}_\mathrm{ext}$ is the in-plane component of the external magnetic field applied to the array.

In the complex representation,
\begin{equation*}
h_0e^{i\Phi}=h_{x,\mathrm{ext}}+ih_{y,\mathrm{ext}}
 +\mathcal{L}_{xx}^{(1)}\frac{\langle{M_x^{(1)}}\rangle}{a^3}
+i\mathcal{L}_{yy}^{(1)}\frac{\langle{M_y^{(1)}}\rangle}{a^3}
 +\mathcal{L}_{xx}^{(2)}\frac{\langle{M_x^{(2)}}\rangle}{a^3}
+i\mathcal{L}_{yy}^{(2)}\frac{\langle{M_y^{(2)}}\rangle}{a^3}\,.
\end{equation*}
Here and hereafter, for the brevity of notation, we omit the subscripts for $Z_1$ and $\kappa_2$ in (\ref{eq:OA11}) and assign these order parameters to sublattice $1$; for sublattice $2$, we introduce notations $Y$ and $\varkappa$, respectively.
With $\langle{M_x^{(1)}}\rangle=M\sin\vartheta_\ast\mathrm{Re}{Z} =M\sin\vartheta_\ast\frac{Z+Z^\ast}{2}$, $\langle{M_y^{(1)}}\rangle=M\sin\vartheta_\ast\mathrm{Im}{Z} =M\sin\vartheta_\ast\frac{Z-Z^\ast}{2i}$,
$\langle{M_x^{(2)}}\rangle=M\sin\vartheta_\ast\frac{Y+Y^\ast}{2}$, $\langle{M_y^{(2)}}\rangle=M\sin\vartheta_\ast\frac{Y-Y^\ast}{2i}$,
one can write down $\mathcal{H}$ for $h_0e^{i\Phi}$ of sublattice $1$:
\begin{equation}\label{eq:302}
\mathcal{H}_1=\mathcal{H}_\mathrm{ext} +\mathcal{H}_{+}^{(1)}Z +\mathcal{H}_{-}^{(1)}Z^\ast  +\mathcal{H}_{+}^{(2)}Y +\mathcal{H}_{-}^{(2)}Y^\ast,
\end{equation}
where, as one can see from the comparison of (\ref{eq:OA7}) to (\ref{eq:207}),
\begin{equation*}
\mathcal{H}_\mathrm{ext}\equiv\frac{\gamma(h_{x,\mathrm{ext}}+ih_{y,\mathrm{ext}})} {2\eta\sin\vartheta_\ast},\qquad
\mathcal{H}_{\pm}^{(1)}\equiv\frac{\gamma M}{4\eta a^3}(\mathcal{L}_{xx}^{(1)} \pm \mathcal{L}_{yy}^{(1)}),\qquad
\mathcal{H}_{\pm}^{(2)}\equiv\frac{\gamma M}{4\eta a^3}(\mathcal{L}_{xx}^{(2)} \pm \mathcal{L}_{yy}^{(2)}).
\end{equation*}
From (\ref{eq:301}), one can find that
\begin{equation*}
\mathcal{H}_{-}^{(2)}=-\mathcal{H}_{-}^{(1)}\equiv\mathcal{H}_{-}
=5.56660...\frac{\gamma M}{4\eta a^3}.
\end{equation*}
It is also convenient to introduce
\begin{equation*}
\mathcal{H}_{+}^\mathrm{fer}\equiv\mathcal{H}_{+}^{(1)}+\mathcal{H}_{+}^{(2)} =4.52180...\frac{\gamma M}{2\eta a^3} \quad\mbox{ and }\quad
\mathcal{H}_{+}^\mathrm{af}\equiv\mathcal{H}_{+}^{(2)}-\mathcal{H}_{+}^{(1)} =0.46273...\frac{\gamma M}{2\eta a^3}.
\end{equation*}
Hence, expression~(\ref{eq:302}) can be rewritten in a shorter form:
\begin{equation}\label{eq:303}
\mathcal{H}_1=\mathcal{H}_\mathrm{ext} +\mathcal{H}_{+}^\mathrm{fer}\frac{Z+Y}{2} -\mathcal{H}_{+}^\mathrm{af}\frac{Z-Y}{2} -\mathcal{H}_{-}(Z^\ast-Y^\ast)\,,
\end{equation}
Similarly, one can write down $\mathcal{H}$ for sublattice $2$:
\begin{equation}\label{eq:304}
\mathcal{H}_2=\mathcal{H}_\mathrm{ext} +\mathcal{H}_{+}^\mathrm{fer}\frac{Z+Y}{2} +\mathcal{H}_{+}^\mathrm{af}\frac{Z-Y}{2} +\mathcal{H}_{-}(Z^\ast-Y^\ast)\,.
\end{equation}

\subsection{Dynamics of mean magnetization of sublattices}
With the magnetic field terms~(\ref{eq:302}) and (\ref{eq:303}), one can employ two-cumulant reduction model~(\ref{eq:OA11}) for the description of the dynamics of order parameters and, thus, macroscopic magnetization. According to equation~(\ref{eq:207}), $\omega=0$ in (\ref{eq:OA11}), and one obtains
\begin{align}
\dot{Z}&=\mathcal{H}_1-\mathcal{H}_1^\ast(Z^2+\kappa)-DZ\,, \label{eq:305} \\
\dot{\kappa}&=-4\mathcal{H}_1^\ast Z\kappa -D(4\kappa+2Z^2)\,, \label{eq:306} \\
\dot{Y}&=\mathcal{H}_2-\mathcal{H}_2^\ast(Y^2+\varkappa)-DY\,, \label{eq:307} \\
\dot{\varkappa}&=-4\mathcal{H}_2^\ast Y\varkappa -D(4\varkappa+2Y^2)\,. \label{eq:308}
\end{align}
In terms of the sublattice order parameters $Z$ and $Y$, the macroscopic in-plane magnetization of the system reads
\begin{equation}\label{eq:309}
\mu_x+i\mu_y=\mu_\ast\frac{Z+Y}{2}\,,
\end{equation}
where the saturation value
\begin{equation*}
\mu_\ast=\frac{M\sin\vartheta_\ast}{a^2}\,,
\end{equation*}
and the parameter of the antiferromagnetic order is
\begin{equation}\label{eq:310}
A\equiv\frac{Z-Y}{2}\,.
\end{equation}

For a time-independent external field $\mathcal{H}_\mathrm{ext}$, equations~(\ref{eq:305})--(\ref{eq:308}) with (\ref{eq:303})--(\ref{eq:304}) yield a dynamics of relaxation to the time-independent states. The algebraic equation system $\dot{Z}=\dot{\kappa}=\dot{Y}=\dot{\varkappa}=0$ is a nonlinear system of high order and its analytical solving is generally not possible. Meanwhile, one can obtain the stable time-independent solutions by the direct numerical simulation of the low-dimensional macroscopic model rigorously derived from the first principles~\footnote{Low-dimensional compared to the infinitely dimensional original system with $N\to\infty$.}.

\subsubsection{Ferromagnetic state}
For ferromagnetic states, $Z=Y$, the complex conjugate terms in (\ref{eq:303}) and (\ref{eq:304}) vanish and $\mathcal{H}_1=\mathcal{H}_2=\mathcal{H}_\mathrm{ext}+\mathcal{H}_{+}^\mathrm{fer}Z$. In this case, equation system~(\ref{eq:305})--(\ref{eq:306}) is equivalent to (\ref{eq:307})--(\ref{eq:308}),
\begin{align}
\dot{Z}&=\mathcal{H}_\mathrm{ext}+\mathcal{H}_{+}^\mathrm{fer}Z-(\mathcal{H}_\mathrm{ext}^\ast+\mathcal{H}_{+}^\mathrm{fer}Z^\ast)(Z^2+\kappa)-DZ\,, \label{eq:311} \\
\dot{\kappa}&=-4(\mathcal{H}_\mathrm{ext}^\ast +\mathcal{H}_{+}^\mathrm{fer}Z^\ast)Z\kappa -D(4\kappa+2Z^2)\,, \label{eq:312}
\end{align}
and becomes invariant to the transform $(\mathcal{H}_\mathrm{ext},Z,\kappa)\rightarrow(\mathcal{H}_\mathrm{ext}e^{i\phi},Ze^{i\phi},\kappa e^{i2\phi})$ with arbitrary $\phi$. From the physical point of view, this means that the system's in-plane dynamics becomes isotropic, as it should be for ferromagnetic states on a square lattice.

In polar coordinates, one can write for the applied field $\mathcal{H}_\mathrm{ext}=\mathcal{H}_{0,\mathrm{ext}}e^{i\phi}$ and for the order parameters $Z=R\,e^{i(\psi_1+\phi)}$ and $\kappa=\rho e^{i(\psi_2+2\phi)}$, where $\mathcal{H}_{0,\mathrm{ext}}$, $R$, and $\rho$ are real. Equation system~(\ref{eq:311})--(\ref{eq:312}) takes the form
\begin{align}
\dot{R}&=\mathcal{H}_{0,\mathrm{ext}}\big[(1-R^2)\cos\psi_1-\rho\cos(\psi_2-\psi_1)\big]+\mathcal{H}_{+}^\mathrm{fer}R\big[1-R^2-\rho\cos(\psi_2-2\psi_1)\big]-DR\,,
\label{eq:313}
\\
\dot\psi_1&=\mathcal{H}_{0,\mathrm{ext}}\big[-(1/R+R)\sin\psi_1 -(\rho/R)\sin(\psi_2-\psi_1)\big] -\mathcal{H}_{+}^\mathrm{fer}\rho\sin(\psi_2-2\psi_1)\,,
\label{eq:314}
\\
\dot{\rho}&=-4(\mathcal{H}_{0,\mathrm{ext}}\cos\psi_1 +\mathcal{H}_{+}^\mathrm{fer}R)R\rho -D\big[4\rho+2R^2\cos(2\psi_1-\psi_2)\big]\,,
\label{eq:315}
\\
\dot\psi_2&=-4\mathcal{H}_{0,\mathrm{ext}}R\sin\psi_1 -2D(R^2/\rho)\sin(2\psi_1-\psi_2)\,.
\label{eq:316}
\end{align}
A thorough examination of equations~(\ref{eq:314}) and (\ref{eq:316}) reveals that these equations have a single attracting manifold $\psi_2=2\psi_1=0$, which corresponds to the alignment of the macroscopic magnetization along the applied magnetic field. On this attracting manifold, equations~(\ref{eq:313}) and (\ref{eq:315}) simplify to
\begin{align}
\dot{R}&=(\mathcal{H}_{0,\mathrm{ext}}+\mathcal{H}_{+}^\mathrm{fer}R)(1-R^2-\rho)-DR\,,
\label{eq:317}
\\
\dot{\rho}&=-4(\mathcal{H}_{0,\mathrm{ext}} +\mathcal{H}_{+}^\mathrm{fer}R)R\rho -D(4\rho+2R^2)\,.
\label{eq:318}
\end{align}

For $\mathcal{H}_\mathrm{ext}=0$, equation system~(\ref{eq:317})--(\ref{eq:318}) possesses time-independent nontrivial solution
\begin{align}
R^2=\frac{1}{2}-\frac{3D}{4\mathcal{H}_{+}^\mathrm{fer}} +\sqrt{\frac14+\frac{D}{4\mathcal{H}_{+}^\mathrm{fer}} -\frac{7D^2}{(4\mathcal{H}_{+}^\mathrm{fer})^2}}\,,
\qquad \rho=-\frac{DR^2}{2D+2\mathcal{H}_{+}^\mathrm{fer}R^2}\,, \label{eq:319}
\end{align}
this solution is always stable and exists for $D<D_\mathrm{fer}=\mathcal{H}_{+}^\mathrm{fer}$. Although solution~(\ref{eq:319}) is globally attracting in the dynamical system (\ref{eq:313})--(\ref{eq:316}), it is globally attracting only on the manifold of the ferromagnetic states; beyond this manifold, competing stable antiferromagnetic states can exit. It is also important, that solution (\ref{eq:319}) is an approximation, where we neglect $\kappa_3$.
However, near the critical point $D=D_\mathrm{fer}$, where this solution tends to $0$, equation chain (\ref{eq:OA10}) yields $\kappa_n\propto Z_1^n$, i.e., the relative inaccuracy of approximation $\kappa_3=0$ tends to $0$ as $Z_1$ tends to zero.
Hence, the critical threshold
\begin{equation*}
D_\mathrm{fer}=\mathcal{H}_{+}^\mathrm{fer}
\end{equation*}
is exact and, employing (\ref{eq:defD}), one can write down the critical temperature for the ferromagnetic state in the absence of an applied in-plane field:
\begin{equation}\label{eq:320}
kT_\mathrm{fer}=\frac{4.52180...\,M^2\sin^2\vartheta_\ast}{2a^3}\,.
\end{equation}

With time-independent ferromagnetic solutions, one can test the accuracy of the two-cumulant reduction. For a constant $\mathcal{H}$ and $\omega=0$, the probability density of angles $\varphi$ governed by Fokker--Planck equation~(\ref{eq:OA8}) tends to the von Mises distribution (see~\cite{Bertini-etal-2010,Goldobin-etal-2018} for details):
\begin{equation*}
w(\varphi)=\frac{\exp\left(\frac{2|\mathcal{H}|}{D}\cos(\varphi-\arg\mathcal{H})\right)}{2\pi I_0(2|\mathcal{H}|/D)}\,,
\end{equation*}
where $I_n(\cdot)$ is the $n$-th order modified Bessel function of the first kind.
With the Jacobi--Anger expansion $e^{a\cos(\varphi-\psi)}=\sum_{n=-\infty}^{+\infty}I_n(a)e^{in(\varphi-\psi)}$, one can calculate $Z=\int_{0}^{2\pi}w(\varphi)e^{i\varphi}\mathrm{d}\varphi =\frac{I_1(2|\mathcal{H}|/D)}{I_0(2|\mathcal{H}|/D)}\frac{\mathcal{H}}{|\mathcal{H}|}$.
Hence, the ferromagnetic state is given by the solution of the following self-consistency problem for $R$\,:
\begin{equation}\label{eq:321}
R=\frac{I_1\big[2(\mathcal{H}_{0,\mathrm{ext}}+\mathcal{H}_{+}^\mathrm{fer}R)/D\big]} {I_0\big[2(\mathcal{H}_{0,\mathrm{ext}}+\mathcal{H}_{+}^\mathrm{fer}R)/D\big]}\,.
\end{equation}
One can compare the numeric solution of equation system (\ref{eq:317})--(\ref{eq:318}) to the numeric solution of self-consistency equation (\ref{eq:321}) in order to test the accuracy of the two-cumulant reduction~(\ref{eq:305})--(\ref{eq:308}).
This comparison shows that, in the physically meaningful range of parameters, the relative inaccuracy of the value of $R$ calculated with the two-cumulant reduction does not exceed $5\%$ and is typically $\sim1\%$.

The analog of the exact self-consistency equation~(\ref{eq:321}) for antiferromagnetic states is enormously lengthy and, more importantly, does not allow one to examine time-dependent regimes or even the stability of the solutions.

Equation (\ref{eq:317}) suggests the reference value for the in-plane external magnetic field
\begin{equation}\label{eq:322}
h_\ast=\frac{2\eta\sin\vartheta_\ast}{\gamma}\mathcal{H}_{+}^\mathrm{fer} =\frac{4.52180...\,M\sin\vartheta_\ast}{a^3}\,;
\end{equation}
for this field, $\mathcal{H}_{0,\mathrm{ext}}=\mathcal{H}_{+}^\mathrm{fer}$.

\subsubsection{Antiferromagnetic state for $\mathcal{H}_\mathrm{ext}=0$}
For the unperturbed antiferromagnetic states, which can be possible at $\mathcal{H}_\mathrm{ext}=0$, one can write $Z=-Y=A$, $\varkappa=\kappa$, and $\mathcal{H}_1=-\mathcal{H}_2=-\mathcal{H}_{+}^\mathrm{af}A -2\mathcal{H}_{-}A^\ast$; therefore, equation system~(\ref{eq:305})--(\ref{eq:308}) simplifies to
\begin{align}
\dot{A}&=-\mathcal{H}_{+}^\mathrm{af}A -2\mathcal{H}_{-}A^\ast +(\mathcal{H}_{+}^\mathrm{af}A^\ast +2\mathcal{H}_{-}A)(A^2+\kappa)-DA\,, \label{eq:323} \\
\dot{\kappa}&=4(\mathcal{H}_{+}^\mathrm{af}A^\ast +2\mathcal{H}_{-}A)A\kappa -D(4\kappa+2A^2)\,. \label{eq:324}
\end{align}
In polar coordinates, $A=\mathcal{A}e^{i\psi_1}$ and $\kappa=\rho e^{i\psi_2}$; equation system~(\ref{eq:323})--(\ref{eq:324}) reads
\begin{align}
\dot{\mathcal{A}}&=\mathcal{H}_{+}^\mathrm{af}\mathcal{A}\big[\mathcal{A}^2-1 +\rho\cos(\psi_2-2\psi_1)\big] +2\mathcal{H}_{-}\mathcal{A}\big[(\mathcal{A}^2-1)\cos2\psi_1+\rho\cos\psi_2\big] -D\mathcal{A}\,, \label{eq:325} \\
\dot\psi_1&=\mathcal{H}_{+}^\mathrm{af}\rho\sin(\psi_2-2\psi_1) +2\mathcal{H}_{-}\big[(1+\mathcal{A}^2)\sin2\psi_1+\rho\sin\psi_2\big]\,, \label{eq:326} \\
\dot{\rho}&=4(\mathcal{H}_{+}^\mathrm{af} +2\mathcal{H}_{-}\cos2\psi_1)\mathcal{A}^2\rho -D\big[4\rho+2\mathcal{A}^2\cos(2\psi_1-\psi_2)\big]\,,  \label{eq:327} \\
\dot\psi_2&=8\mathcal{H}_{-}\mathcal{A}^2\sin2\psi_1 -2D(\mathcal{A}^2/\rho)\sin(2\psi_1-\psi_2)\,. \label{eq:328}
\end{align}
A thorough examination of equations~(\ref{eq:326}) and (\ref{eq:328}) reveals that these equations have attracting manifolds $\psi_1=\psi_2/2=\pm\pi/2$, which corresponds to the alignment of sublattice magnetisations along the sublattice stripes (see figure~\ref{fig1}\textit{b}).
This result is also obvious and expected from the physical point of view.
The manifolds $\psi_1=\pm\pi/2$ are physically equivalent; they differ from each other merely by the interchange of sublattices.
On these manifolds, equations~(\ref{eq:325}) and (\ref{eq:327}) acquire a simplified form:
\begin{align}
\dot{\mathcal{A}}&= \big[(2\mathcal{H}_{-}-\mathcal{H}_{+}^\mathrm{af})(1-\mathcal{A}^2-\rho) -D\big]\mathcal{A}\,, \label{eq:329} \\
\dot{\rho}&=-4(2\mathcal{H}_{-}-\mathcal{H}_{+}^\mathrm{af})\mathcal{A}^2\rho -D(4\rho+2\mathcal{A}^2)\,.  \label{eq:330}
\end{align}

For a time-independent solution of (\ref{eq:329})--(\ref{eq:330}), one can find
\begin{align}\label{eq:331}
\mathcal{A}^2=\frac{1}{2}-\frac{3D}{4D_\mathrm{af}} +\sqrt{\frac14+\frac{D}{4D_\mathrm{af}} -\frac{7D^2}{(4D_\mathrm{af})^2}}\,, \qquad \rho=-\frac{DR^2}{2D+2D_\mathrm{af}R^2}\,,
\end{align}
where
\begin{align}\label{eq:332}
D_\mathrm{af}\equiv2\mathcal{H}_{-}-\mathcal{H}_{+}^\mathrm{af}\,.
\end{align}
The antiferromagnetic state~(\ref{eq:331}) exists for $D<D_\mathrm{af}$ and is always globally attracting in the dynamical system~(\ref{eq:325})--(\ref{eq:328}), but beyond the subspace of antiferromagnetic states there can be competing ferromagnetic states. Equations~(\ref{eq:defD}) and (\ref{eq:331}) yield the critical temperature for the antiferromagnetic state:
\begin{equation} \label{eq:333}
kT_\mathrm{af}=\frac{5.10387...\,M^2\sin^2\vartheta_\ast}{2a^3}\,.
\end{equation}

With no external magnetic field applied, the antiferromagnetic states are the minimal energy states.
Therefore, the critical temperature of the strongly-interacting samples in~\cite{Leo-etal-2018} is $T_\mathrm{af}$.

\subsection{Free energy of macroscopic states with two magnetic sublattices  \label{sec:4c}}
The coexistence of different metastable phase states raises the question of observability of these states.
Without the analysis of the formation of domains of different macroscopic phase states, it is neither possible to describe the transitions between metastable phase states nor evaluate the lifetime of metastable states. However, one can calculate the free energy of macroscopically homogeneous states and find which states provide the global minimum of the free energy.

For the calculation of the partition function and thermodynamic potentials we assume the polar angle degree of freedom to be frozen-out and account only for the azimuthal one.
In the thermodynamic limit of infinitely large ensemble, one can consider the mean fields acting on single-domain particles to be constant and calculate the partition function for individual particles subject to these fields.
For the ensemble of magnetic moments~(\ref{eq:207}), the energy of an ensemble state $\mathcal{E}=-\sum_{j=1}^{N}(\vec{M}_j\cdot\vec{h}_j) =-\sum_{j=1}^{N}M\sin\vartheta_\ast|h_j|\cos(\varphi_j-\Phi)$. Hence, the partition function
\begin{equation*}
\Xi=\int...\int\mathrm{d}\varphi_1...\mathrm{d}\varphi_N\exp\left(-\frac{\mathcal{E}}{kT}\right) =\prod_{j=1}^N\int\mathrm{d}\varphi_j\exp\left(\frac{\vec{M}_j\cdot\vec{h}_j}{kT}\right)\,.
\end{equation*}
Free energy
\begin{equation*}
\mathcal{F}=-\frac{\partial}{\partial(\frac{1}{kT})}\ln\Xi\,.
\end{equation*}

For a two-sublattice ensemble
\begin{equation*}
\mathcal{F}=-\frac{N}{2}\frac{\partial}{\partial(\frac{1}{kT})}\ln\Xi_2\,,
\end{equation*}
where
\begin{align}
\Xi_2&=\int\mathrm{d}\varphi^{(1)} \exp\left(\frac{|h_1|M\sin\vartheta_\ast\cos(\varphi^{(1)}-\Phi_1)}{kT}\right) \int\mathrm{d}\varphi^{(2)} \exp\left(\frac{|h_2|M\sin\vartheta_\ast\cos(\varphi^{(2)}-\Phi_2)}{kT}\right) \nonumber \\
&=I_0\left(\frac{|h_1|M\sin\vartheta_\ast}{kT}\right)\, I_0\left(\frac{|h_2|M\sin\vartheta_\ast}{kT}\right)\,, \nonumber
\end{align}
$\vec{h}_1$ and $\vec{h}_2$ are the magnetic fields at sublattice nodes.
Hence,
\begin{align}
\mathcal{F}_1&\equiv\frac{\mathcal{F}}{N}=-\frac{1}{2}\frac{\partial}{\partial(\frac{1}{kT})}\ln\Xi_2\,  \nonumber\\
&
=-\frac{|h_1|M\sin\vartheta_\ast}{2}\frac{I_1\left(\frac{|h_1|M\sin\vartheta_\ast}{kT}\right)} {I_0\left(\frac{|h_1|M\sin\vartheta_\ast}{kT}\right)} -\frac{|h_2|M\sin\vartheta_\ast}{2}\frac{I_1\left(\frac{|h_2|M\sin\vartheta_\ast}{kT}\right)} {I_0\left(\frac{|h_2|M\sin\vartheta_\ast}{kT}\right)}\,. \nonumber
\end{align}
Comparing equation (\ref{eq:OA7}) to (\ref{eq:207}), one can recast $\mathcal{F}_1$ in terms of $\mathcal{H}_1$ and $\mathcal{H}_2$:
\begin{equation*}
\frac{\mathcal{F}_1}{\mathcal{F}_\ast}=-|\mathcal{H}_1|\frac{I_1\left(2|\mathcal{H}_1|/D\right)} {I_0\left(2|\mathcal{H}_1|/D\right)} -|\mathcal{H}_2|\frac{I_1\left(2|\mathcal{H}_2|/D\right)} {I_0\left(2|\mathcal{H}_2|/D\right)}\,,
\end{equation*}
where the reference value of energy
\begin{equation*}
\mathcal{F}_\ast=\frac{M^2\sin^2\vartheta_\ast}{a^3}\,.
\end{equation*}
Similarly to (\ref{eq:321}), for a thermodynamic equilibrium state, one can obtain $|Z|=\frac{I_1\left(2|\mathcal{H}_1|/D\right)}{I_0\left(2|\mathcal{H}_1|/D\right)}$ and $|Y|=\frac{I_1\left(2|\mathcal{H}_2|/D\right)}{I_0\left(2|\mathcal{H}_2|/D\right)}$, and rewrite
\begin{equation}
\frac{\mathcal{F}_1}{\mathcal{F}_\ast}=-|\mathcal{H}_1||Z| -|\mathcal{H}_2||Y|\,.
\label{eq:334}
\end{equation}
With equation (\ref{eq:334}) the free energy of macroscopic states can be evaluated from the results of the numerical simulations of the two-cumulant reduction model (\ref{eq:305})--(\ref{eq:308}) with (\ref{eq:303}), (\ref{eq:304}).

\begin{figure}[!h]
\centering
\textit{(a)}\hspace{-10pt}\includegraphics[height=1.7in]{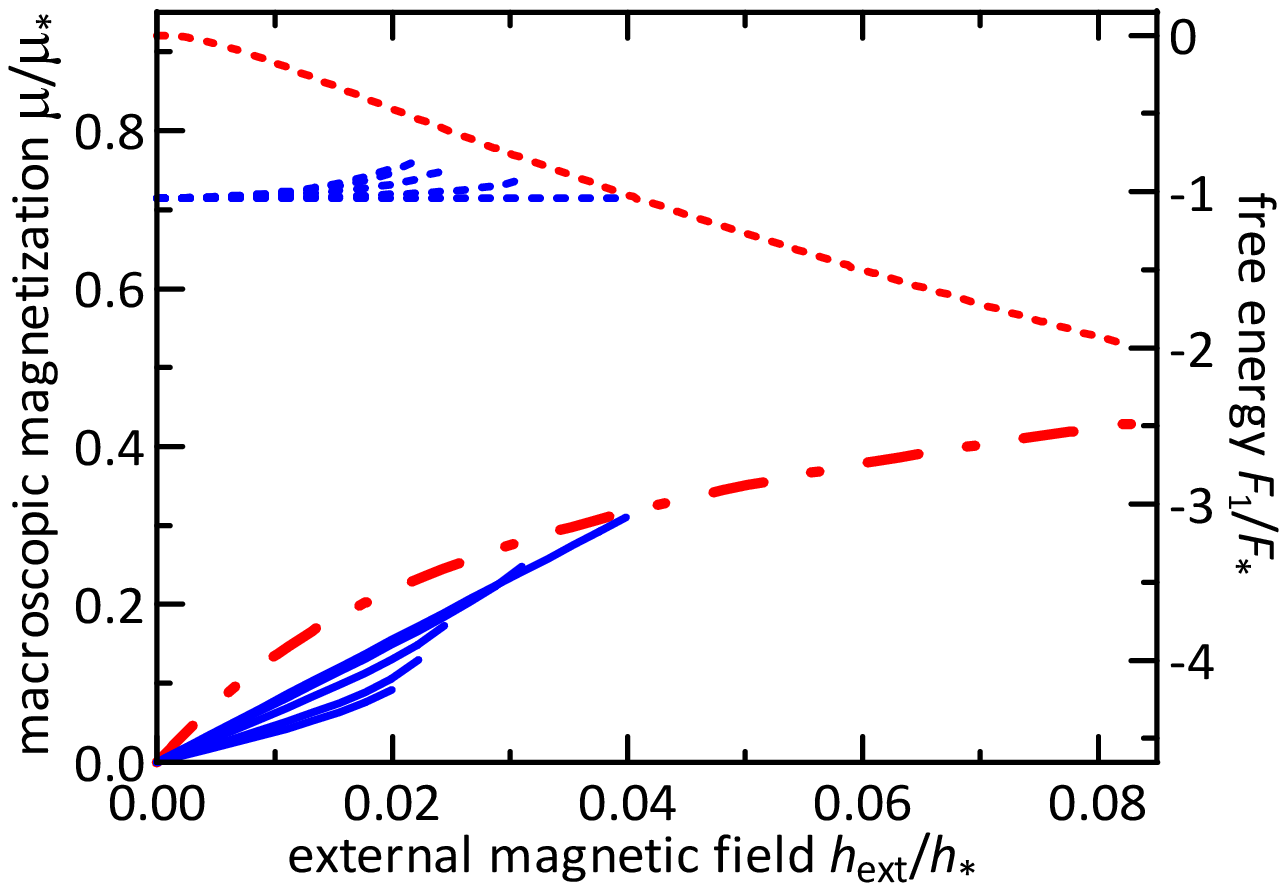}
\quad
\textit{(b)}\hspace{-10pt}\includegraphics[height=1.7in]{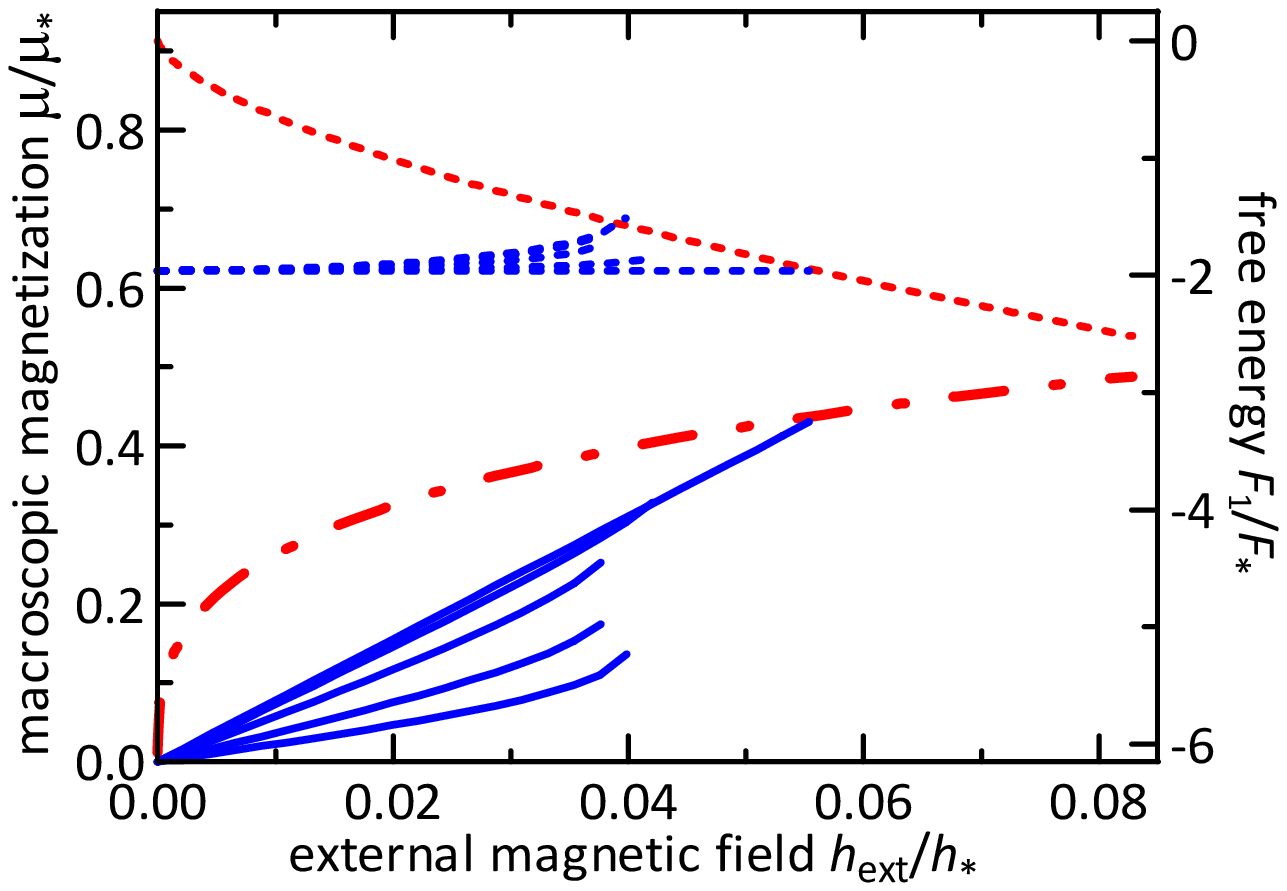}
\\[12pt]
\textit{(c)}\hspace{-10pt}\includegraphics[height=1.7in]{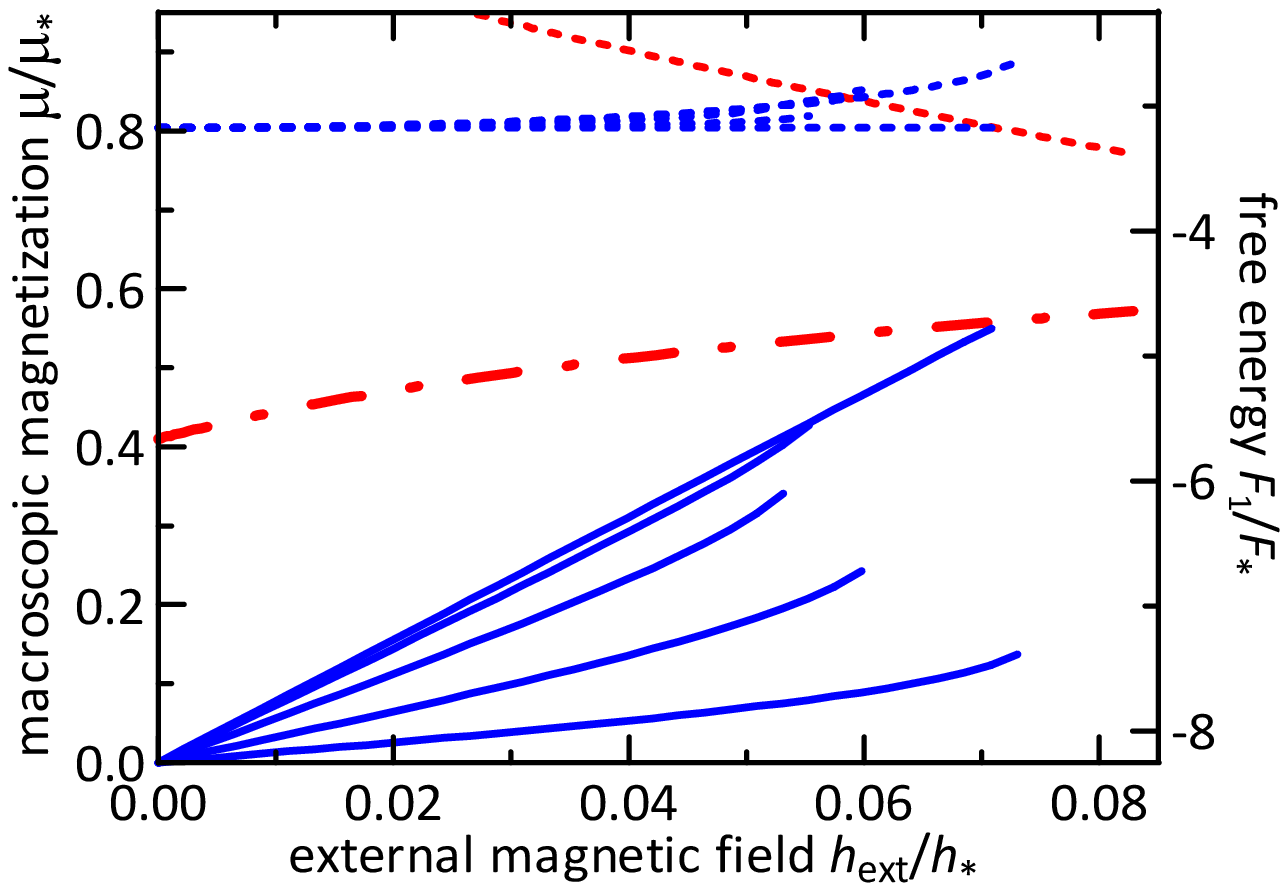}
\quad
\textit{(d)}\hspace{-10pt}\includegraphics[height=1.7in]{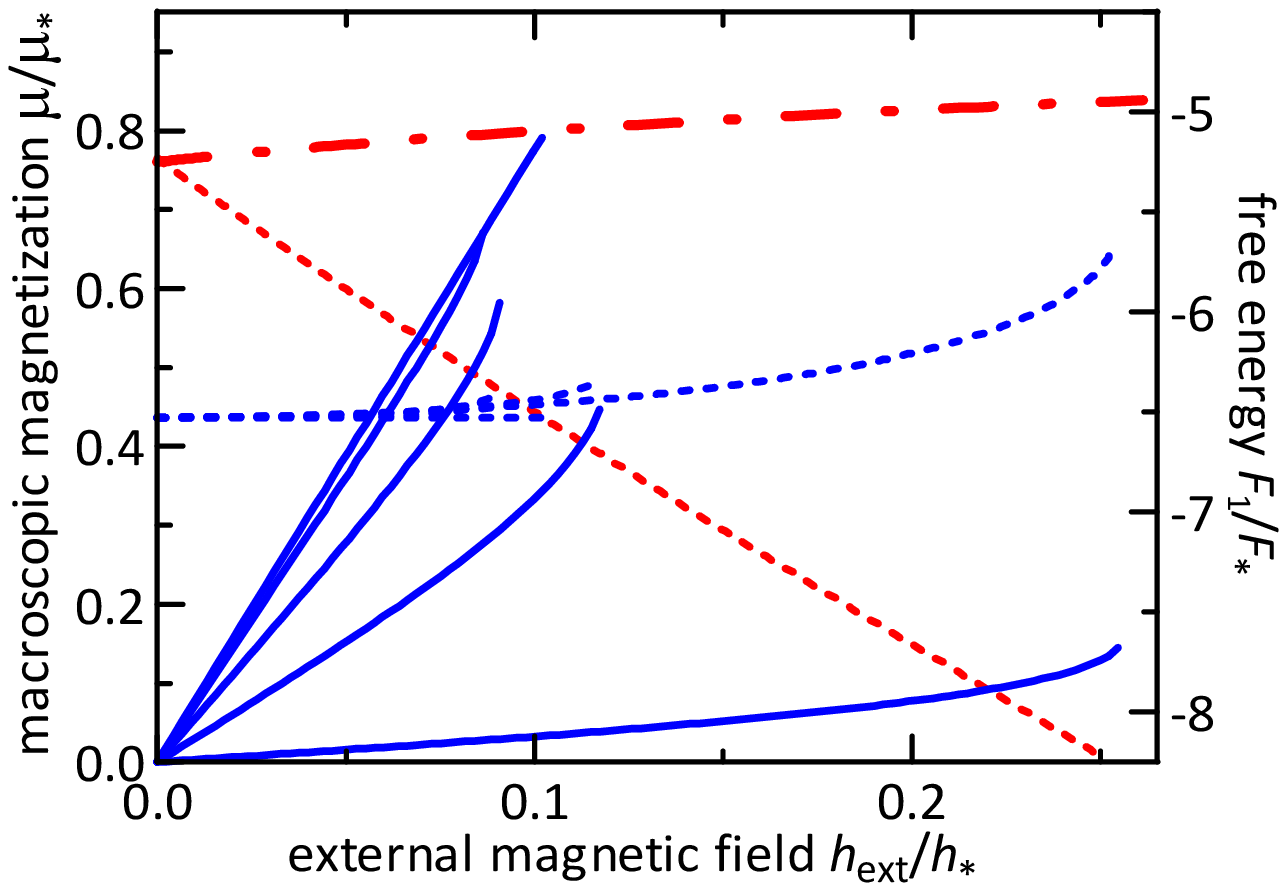}
\caption{For temperature $T=(T_\mathrm{fer}+T_\mathrm{af})/2$~(\textit{a}), $T_\mathrm{fer}$~(\textit{b}), $0.8T_\mathrm{af}$~(\textit{c}), and $0.5T_\mathrm{af}$~(\textit{d}), the macroscopic in-plane magnetization $\mu$ is plotted versus the applied in-plane magnetic field $h_\mathrm{ext}$ with the red dash-dotted curve for the ferromagnetic state and with the blue solid curves for the antiferromagnetic states.
While the former is isotropic, the latter depends on the orientation of the applied field: $\phi=0$, $\pi/8$, $\pi/4$, $3\pi/8$, $\pi/2$ (from top to bottom; see figure~\ref{fig1}\textit{b} for the definition of $\phi$).
The free energy (\ref{eq:334}) of these states is plotted with dotted curves.
The details presented for $T=0.8T_\mathrm{af}$~(\textit{c}) in figure~\ref{fig3} are qualitatively similar for all reported cases.}
\label{fig2}
\end{figure}

\begin{figure}[!h]
\centering
\includegraphics[height=1.7in]{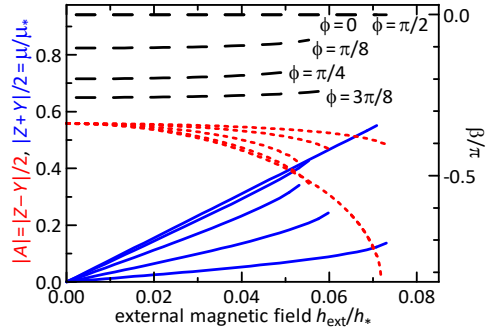}
\caption{For $T=0.8T_\mathrm{af}$~(figure~\ref{fig2}\textit{c}), the macroscopic magnetization $\mu$ is plotted with the blue solid curves for $\phi=0$, $\pi/8$, $\pi/4$, $3\pi/8$, $\pi/2$ (from top to bottom); the antiferromagnetic order parameter $\mathcal{A}$ is plotted with the red dotted curves; the angle $\beta$ between the applied in-plane field $\vec{h}_\mathrm{ext}$ and the macroscopic magnetization $\vec{\mu}$  (see figure~\ref{fig1}\textit{b}) is plotted with the black dashed curves.}
\label{fig3}
\end{figure}

\section{Magnetization and phase transitions in the system subject to an external magnetic field  \label{sec:5}}
In this section we present the results of numerical simulation of the two-cumulant model reduction (\ref{eq:305})--(\ref{eq:308}) with (\ref{eq:303}), (\ref{eq:304}) and interpret them from the viewpoint of macroscopic observations.
Two critical temperatures are important for the system behavior: $T_\mathrm{af}$ (\ref{eq:333}), above which the antiferromagnetic ordering disappears, and $T_\mathrm{fer}\approx0.88596\,T_\mathrm{af}$ (\ref{eq:320}), above which the metastable ferromagnetic phase state becomes impossible for $h_\mathrm{ext}=0$.
The critical temperature reported in experiments with the strongly-interacting samples in~\cite{Leo-etal-2018} should correspond to $T_\mathrm{af}$.
The numerical simulation reveals that, above $T_\mathrm{af}$, only ferromagnetic states are possible in the system subject to the external in-plane magnetic field.

In figures \ref{fig2}\textit{a},\textit{b}, one can see, that, in the temperature range $T_\mathrm{fer}<T<T_\mathrm{af}$, antiferromagnetic states provide the minimum of the free energy; where the antiferromagnetic state exists, the ferromagnetic state is metastable (see dotted curves in figures \ref{fig2}\textit{a},\textit{b}).
The ferromagnetic state is isotropic and, therefore, the magnetic susceptibility of the system (red dash-dotted curve) does not depend on the applied field orientation above the critical field strength, where the antiferromagnetic state disappears.
In figure \ref{fig3}, one can see that, for $\phi\ne0$, the antiferromagnetic order parameter $\mathcal{A}$ (dotted curves) is nonzero at the critical strength of external field $h_\mathrm{ext}$, i.e., the antiferromagnetic state disappears via a first-order phase transition.
Only for $\phi=0$, i.e., the applied field orthogonal to the stripes of magnetic sublattices (figure~\ref{fig1}\textit{b}), the disappearance of the antiferromagnetic state is a second-order phase transition.
In figures \ref{fig2}\textit{a},\textit{b}, one can also see, that with given $h_\mathrm{ext}$ the free energy is minimal for smaller $\phi$.
This is important, since the cases $\phi$ and $\pi/2-\phi$ correspond to the same orientation of the array with respect to the external field, but different arrangement of the magnetic sublattices (in figure \ref{fig1}\textit{b}, the alignment of magnetic sublattices 1 and 2 along the $x$-axes corresponds to the switching from $\phi$ to $\pi/2-\phi$).
Thus, the magnetic sublattices tend to align perpendicularly to the external field and the states with a nearly parallel alignment of magnetic sublattices ($\phi>\pi/4$) become metastable; given enough time, they switch from the case of $\phi$ to the case of $\pi/2-\phi$.

In figure \ref{fig3}, one can see that the angle $\beta$ between the macroscopic magnetization and the external field (see figure~\ref{fig1}\textit{b}) is zero for $\phi=0$ and becomes negative as $\phi$ grows.
The external field deflects the magnetization from the $x$-axis, although for small $\phi$, $\beta\approx-\phi$, meaning the magnetization is almost parallel to the $x$-axes (perpendicular to magnetic sublattices). As the external field becomes stronger the absolute value of $\beta$ decreases, i.e., the magnetization orientation is attracted towards the one of the external field; but the variation of $\beta$ is quite small up to the critical strength of the field and the destruction of the antiferromagnetic state.
The angle $\beta$ monotonously changes with growing $\phi$ until a very small vicinity of $\pi/2$, where $\beta$ fast tends to $0$ and the magnetization becomes again parallel to the applied field for $\phi=\pi/2$.
Recall, however, that the states with $\phi>\pi/4$ are metastable in the discussed temperature range from $T_\mathrm{fer}$ to $T_\mathrm{af}$.

In figures \ref{fig2}\textit{c},\textit{d}, one can see that below $T_\mathrm{fer}$, the metastable antiferromagnetic state can possess a higher free energy than the ferromagnetic state (e.g., for $\phi=\pi/2$ and large $h_\mathrm{ext}$).
The antiferromagnetic state with $\phi\le\pi/4$ is still providing the free energy minimum compared to both the antiferromagnetic state with $\phi>\pi/4$ and the ferromagnetic state.
Below $T\approx0.8T_\mathrm{af}$, the existence domain of the antiferromagnetic state with $\phi$ close to $\pi/2$ expands beyond that of the states with $\phi<\pi/4$.
Although the antiferromagnetic states with $\phi>\pi/4$ can exist for the field strength, where the antiferromagnetic states with $\phi<\pi/4$ are already destroyed, these persisting states possess a higher free energy than the ferromagnetic state. Thus, they are always metastable and there is no additional phase transitions in the system associated with these states.
To summarize, the antiferromagnetic states with $\phi<\pi/4$ are thermodynamically stable as long as they exist; at a critical strength of the external field, the stable antiferromagnetic state is destroyed and a first-order phase transition to the ferromagnetic state occurs in the system.
In the particular case of the external field perfectly orthogonal to the magnetic sublattices (i.e., $\phi=0$), the phase transition is of the second order.

\section{Conclusion  \label{sec:6}}
We have demonstrated the application of the circular cumulant approach \cite{Tyulkina-etal-2018,Tyulkina-etal-2019,Goldobin-Dolmatova-PRR-2019,Goldobin-2019} (a generalization of the Ott--Antonsen theory) to macroscopic description of the collective magnetism phenomena in systems with one principal angular degree of freedom---such as XY spin systems.
The original OA theory~\cite{Ott-Antonsen-2008,Ott-Antonsen-2009} could find only very limited applications for the collective magnetism problems, as it cannot handle the thermal noise and can deal with quite particular types of nonidentities of parameters of individual elements.
On the basis of the circular cumulant approach, one can generalize the OA theory and derive closed equation systems for the dynamics of order parameters in the presence of thermal noise ({\it or} `intrinsic noise') and other violations of the applicability conditions of the original OA theory.

We have derived the equation of the dynamics of the azimuthal angle of individual magnetic moments of single-domain nanoparticles~\cite{Arnalds-etal-2014,Leo-etal-2018} from the first principles.
For the population of such directional elements on a 2D square lattice with the dipole--dipole interaction, we have obtained a closed set of equations for the two leading order parameters (circular cumulants) for each magnetic sublattice: equations~(\ref{eq:305})--(\ref{eq:308}) with (\ref{eq:303}) and (\ref{eq:304}).
With these parameters, one can track the dynamics of the macroscopic magnetization and the parameter of antiferromagnetic order.
Numerical simulation and analytical examination of the cumulant equations reveal a comprehensive picture of the system response to a static external magnetic field and possible phase transitions due to the change in temperature or in the applied field.

The principal physical assumption of our mathematical model is that we adopt the mean-field approximation.
Hence, the formation of domains of different macroscopic states is excluded from our analysis, and we cannot describe the transitions between different phase states in detail.
Nonetheless, we have calculated the free energy of macroscopic states and identified which of them are thermodynamically stable and which are metastable.

Noteworthy, the Ott--Antonsen theory can be used for the description of the macroscopic dynamics of ensembles with nonlocal interaction, as one can see for 1D systems in \cite{Laing-2009,Bordugov-Pikovsky-Rosenblum-2010,Omelchenko-etal-2014,Smirnov-etal-2017,Smirnov-etal-2018,Omelchenko-2019} and for 2D systems in~\cite{Laing-2017,Omelchenko-2018}.
The formation of domains can be described on the basis of circular cumulant reductions in a similar way, but this issue is beyond the scope of our paper.
Moreover, the in-plane magnetic anisotropy of the XY spin systems, which is important for the weakly-interacting samples~\cite{Leo-etal-2018}, results in an additional $e^{i2\varphi}$-term in equation (\ref{eq:OA7}).
Such a term violates the form of equations required for the original Ott--Antonsen theory, but can be naturally handled within the framework of the circular cumulant approach~\cite{Tyulkina-etal-2018,Goldobin-Dolmatova-2019b}.

\aucontribute{IVT, DSG, and LSK derived the stochastic equation~(\ref{eq:207}) and developed the circular cumulant representation of the two-sublattice mean-field model. IVT conducted the numerical simulation of the two-cumulant model; DSG performed its analytical examination. YLR, ISP, and DSG developed the models and interpreted the mathematical results. DSG, ISP, and YLR conceived and designed the study; DSG and YLR drafted the manuscript. All authors read and approved the manuscript.}

\competing{The authors declare that they have no competing interests.}

\funding{The work of IVT, DSG, and LSK was supported by a joint RSF--DFG project (Russian Science Foundation grant no.\ 19-42-04120).}


\end{document}